# Collective Alignment of Cells in Planar Cell Polarity: Insights from a Spin Model

Talagadadeevi Nirupamateja[1], Siva Sai Himakar Sreerangam[1, $], Biplab Bose[1]

[1]*Department of Biosciences and Bioengineering,*

*Indian Institute of Technology Guwahati, Guwahati, India 781039*

E-mail: *biplabbose@iitg.ac.in*

[$]*Current Affiliation: American Express, Gurugram, India*

In metazoans, cells collectively polarize and align along the tissue plane. This phenomenon is called Planar cell polarity (PCP). Polarization means asymmetric segregation of molecules and sub-cellular structures within a cell. In PCP, cells collectively align in a particular direction along the tissue plane through identical polarization. PCP in the *Drosophila* wing requires local cell-cell interactions in the presence of some global cue. We used a lattice-based equilibrium model and investigated the collective alignment of cells through local interactions and a global cue. This system undergoes a percolation transition and belongs to the universality class of 2D random percolation. We show that the local interaction should be beyond a threshold to trigger system-level coordinated polarization of cells. Under this condition, even a weak global cue can align all cells in the correct direction. With strong local interactions, this system is robust against local aberrations in global signaling, and collective alignment of cells is achieved even with a transient global signal.

## I. Introduction

Most cells in a multicellular organism are polarized. By polarization of a cell, we mean asymmetric distribution of molecules and sub-cellular structures in a cell [1]. For example, epithelial cells lining our small intestine show apico-basal polarity. Specialized structures and molecules required for the absorption of nutrients are present only on the apical side of these cells that face the intestinal lumen.

In certain tissues, an ensemble of cells collectively polarizes and aligns in a particular direction along the tissue plane. This polarization is orthogonal to apico-basal polarity. This phenomenon is called planar cell polarity (PCP) [2,3]. PCP plays a vital role in the development and physiology of metazoans [4,5]. In mammals, PCP is involved in the development of the kidney [6], spermatogenesis [7], development of neural tube [8], and orientation of sensory hair cells in the inner ear [9].

The wing of *Drosophila* is one of the best-characterized model systems for investigating PCP.

*Drosophila* has hair-like structures called Trichomes on its wing. These trichomes are oriented in a particular direction as they protrude from the distal side of epithelial cells on the wing [10]. The directional arrangement of trichomes across all cells in the wing emerges from the collective polarization of cells [10].

Polarisation of cells in the *Drosophila* wing involves two molecular processes – a) local interaction, which includes the molecular interactions within a cell and between neighboring cells, and b) global signaling that specifies the proximal-distal direction [3].

Local interaction is achieved through the core PCP pathway [3,11]. The key molecules in this pathway are Diego (Dgo), Dishevelled (Dsh), Frizzled (Fz), Flamingo (Fmi), Prickle (Pk), and Van Gogh (Vang). A cell gets polarised through the gradual asymmetric segregation of these proteins [3]. Vang and Pk form a complex that accumulates on the proximal edges of a cell. The Fz-Dsh-Dgo complex accumulates on the distal side of the cell.

Several intracellular processes promote the mutual segregation of these complexes inside a cell [3,11]. Additionally, the Vang-pk complex on the edges of one cell interacts favorably with the Fz-Dsh-Dgo complex on the edges of its neighbors [3]. This intercellular interaction acts like feedback and leads to the asymmetric segregation of these complexes within a cell.

The local interactions through the core PCP pathway will polarize an ensemble of cells in any direction. However, a tissue-level global cue is needed to specify the proximal-distal direction and promote the collective polarization of cells in that direction only. Over the years, several candidates for the global cue have been proposed, like the gradient in the expression of Fourjointed and Dachsous, secreted ligands of the Wnt pathway, and mechanical tension during wing morphogenesis [12]. However, the identity of the global cue in *Drosophila* wing PCP is still debated [13].

Earlier, we used a lattice-based model to capture the essential features of the collective polarization and alignment of cells through local interactions [14]. This model did not consider any global cue. In this model, the protein complexes are represented by spins (+1 and −1). These spins are distributed along the edges and move from one edge to another. Inside the cell, opposite spins tend to exclude each other mutually. However, opposite spins on the edges of neighboring cells stabilize each other.

We investigated the equilibrium behavior of this model. It was observed that local interactions between spins lead to the emergence of large clusters of cells aligned in one direction. Interestingly, the formation of clusters of aligned cells showed a percolation transition like the simple 2D random percolation [14].

No global cue was considered in this spin model. As a result, collective alignment emerged in multiple possible directions. A similar phenomenon is expected even in real cells undergoing PCP. The vectorial polarization of cells along the proximal-distal axis can not be achieved through local interactions. A tissue-level global cue (transient or prolonged) must guide the polarization process.

In the present work, we modified the spin-based PCP model and included a global cue. This model considers a uniform field that acts on the spins and promotes the movement of spins in a particular direction. This uniform field is equivalent to the (currently debated) tissue-level global cue in PCP.

We investigated the equilibrium behavior of this modified model. In this model, clusters of cells aligned in a particular direction, specified by the global cue, emerge, and the system undergoes a percolation transition. We show that this system belongs to the universality class of simple 2D random percolation.

Further, we investigated the importance of local interactions and the global cue in PCP. We show that local interactions are essential to make the system robust against local disruptions of global signaling. Further, a strong local interaction above a threshold primes cells for collective polarization, facilitating even a weak global signal to trigger alignment in a particular direction.

## II. The model

In our earlier work [14], we investigated a spin model based on the work of Strandkvist [15]. In the present work, we modified the model to include a global cue. The epithelial cells in the *Drosophila* wing are roughly hexagonal. In the model, hexagonal cells are arranged on a two-dimensional triangular lattice with periodic boundary conditions. Each cell has six neighbors. Every edge of a cell has an associated spin ($S$), where $S = -1, 0, +1$. Each cell has two copies of these three spins. $S = -1$ and $+1$ represent two protein complexes involved in the PCP. Whereas $S = 0$ means the absence of any protein complex.

In *Drosophila* wing PCP, two protein complexes accumulate on opposite sides within a cell. Therefore, a configuration with $S = -1$ and $+1$ on opposite edges of a cell is favored. However, opposite protein complexes between neighboring cells have favorable interactions. So, a configuration with opposite spins ($S = -1$ and $+1$) on the adjacent edges of neighboring cells is favored over other configurations. These favored configurations are explained in Fig. 1a.

Interactions of spins within a cell and with the neighboring cells represent local interactions. The Hamiltonian for this local interaction is:

$$h = m_1 \sum_{\substack{\text{all n} \\ \text{pairs}}} S_{\alpha,p} \times S_{\beta,q} + m_2 \sum_{\alpha=1}^{N} \sum_{\substack{\text{all 6} \\ \text{pairs}}} S_{\alpha,a} \times S_{\alpha,b} \quad (1)$$

$\underbrace{\phantom{m_1 \sum S_{\alpha,p} \times S_{\beta,q}}}_{\text{cell-cell interactions}}$ $\underbrace{\phantom{m_2 \sum \sum S_{\alpha,a} \times S_{\alpha,b}}}_{\text{within a cell interactions}}$

Here, $m_1 > 0$ and $m_2 < 0$. These two dimensionless parameters represent the relative strengths of spin-spin interactions over the noise in the system.

$m_1$ is the parameter for interaction between spins on neighboring cells. In Eq. (1), α and β are neighboring cells such that the p$^{th}$ edge of cell α is adjacent to q$^{th}$ edge of cell β. $S_{\alpha,p}$ is the spin of the p$^{th}$ edge of cell α. $S_{\beta,q}$ is the spin of the q$^{th}$ edge of cell β.

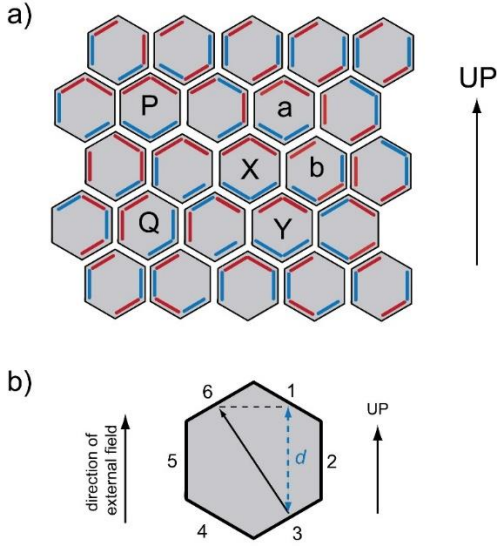

**Figure 1.** Arrangement of cells and their spins. a) Each grey hexagon is a cell. Edges of each cell have associated spins: S = +1 (red bar), S = −1 (blue bar), and S = 0 (no bar). Opposite spins inside a cell should segregate to opposite edges, and adjacent cells should have favorable spin-spin interactions. Following the global cue, cells should align in the correct direction (UP), shown by the vertical arrow. A cell is correctly aligned if the +1 spins (red bars) are on the upper two edges, the −1 spins (blue bars) are on the opposite edges, and all the spin-spin interactions with the adjacent cells are favorable. Here, cells P, X, and Y are correctly aligned. Cells Q, a, and b are not correctly aligned. b) Six edges of a cell are marked. Suppose, at an MC step, the spin from edge 3 of the selected cell is swapped with the spin of edge 6. Spin at edge 3 is moving a vertical distance d. As it moves in the direction of the external field, the sign of this distance is considered positive. Spin at edge 6 is moving the same vertical distance but with a negative sign.

For a lattice of size $L$, the total number of cells $N = L^2$ and the number of interaction pairs between neighbors $n = 3N$.

$m_2$ is the parameter for interaction between adjacent spins within a cell. In Eq. (1), $S_{\alpha,a}$ and $S_{\alpha,b}$ are the spins of two adjacent edges of the cell $\alpha$.

We used Monte Carlo (MC) simulations [16] to understand the equilibrium behavior of this system. Initially, spins of edges in every cell are assigned randomly. At every MC step, we select a cell randomly and then randomly select two edges of that cell. Subsequently, we swap the spins associated with these two edges. Swapping spins is equivalent to changing the position of protein complexes within a cell.

Swapping of spins leads to a change in energy ($\Delta h$) of the system due to local interactions. We calculate $\Delta h$ using Eq. (1).

Using this model, we showed earlier that at equilibrium, cells on the lattice can align in any of the six possible directions [14]. Achieving alignment of all the cells in a specific direction requires a global cue. Our updated model considers a global cue that directs the cells to align in the UP direction, as explained in Fig. 1a and 1b. In *Drosophila*, cells get polarized along the proximal-distal axis. Therefore, one can consider that the UP direction in our model is pointing to the distal side of the wing.

We represent the global cue by a uniform external field interacting with the cell's spins. The spin swap at every MC step leads to a change in energy due to the interaction of spins with this external field. This change in energy is given by:

$$\Delta g = m_3 \left( d_a S_{\alpha,a} + d_b S_{\alpha,b} \right) \quad (2)$$

Here, $S_{\alpha,a}$ and $S_{\alpha,b}$ are the spins swapped. $m_3 < 0$ is a dimensionless parameter that represents the strength of the external field. $|d_a| = |d_b| = d$ is the unitless vertical distance between the edges involved in the spin swap. This distance is calculated considering a cell as a hexagon (see Fig. 1b). $d_a$ and $d_b$ have opposite signs. The sign is positive if the spin is moving in the direction of the field (i.e., UP direction). Otherwise, the sign is negative. Therefore, the movement of +1 spins in the UP direction is favored and the opposite for the −1 spins.

At every MC step, we swap the spins of two randomly selected edges and calculate the change in total energy ($\Delta \varepsilon$) due to local interactions and the global cue:

$$\Delta \varepsilon = \Delta h + \Delta g \quad (3)$$

The spin swap is accepted with the probability $P = \min\left(e^{-\Delta \varepsilon}, 1\right)$.

Three lattice sizes $L$ = 128, 256, and 512, with the periodic boundary condition, were used for simulations. A large number of samples were collected at equilibrium, and different parameters were calculated from those sample configurations.

# III. Results

## A. Collective alignment of cells in the right direction

In PCP, cells get polarized with the segregation of molecules along the proximal-distal axis. So, we say that these cells are aligned in the proximal-distal axis. In this work, we investigate the alignment of an ensemble of cells in a particular direction and the formation of clusters of aligned cells. In our model, a cell is correctly aligned (or aligned UP) if it fulfills three criteria− a) all spins are favorably segregated within the cell, b) all spin-spin interactions with adjacent cells are favorable, and c) +1 spins are on the upper edges (edges 1 and 6) of the cell. Following this definition, cells *P*, *X*, and *Y* in Fig. 1a are correctly aligned.

Correctly aligned, adjacent cells form a cluster of aligned cells. In Fig. 1a, cells *X* and *Y* form a cluster of correctly aligned cells. The size of this cluster is two. The minimum size of a cluster is one, and cell *P* in Fig. 1a is a cluster of size one.

Our model has three control parameters, $m_1$, $m_2$ and $m_3$. We considered $|m_1| = |m_2| = m$ and investigated the model's behavior for various values of $m_3$ and $m$. First, we investigated the alignment of cells in the correct direction (i.e., UP direction).

Fig. 2a shows the system's behavior for various values of $m_3$, for $m = 0.5$. The fraction of cells aligned in the UP direction increases smoothly with the increase of the strength of the global cue $|m_3|$. For $|m_3| \geq 7$, almost all the cells are aligned in the correct direction. This behavior was observed for all three lattice sizes used.

We performed combinatorial simulations with $0 \leq m \leq 4$ and $0 \leq |m_3| \leq 10$. The data for the fraction of cells aligned UP ( $f_U$ ) for different combinations of $m$ and $|m_3|$ is shown in Fig 2b. The dark blue region in the plot represents complete alignment in the UP direction (i.e., $f_U$ close to 1). It is evident that a large parameter space is available to the system to achieve the correct alignment of all the cells across the lattice. Note that complete alignment in the correct direction can happen even without local interactions ( $m = 0$ ), albeit for a high value of $|m_3|$. However, correct alignment is not possible without the global cue ( $m_3 = 0$ ).

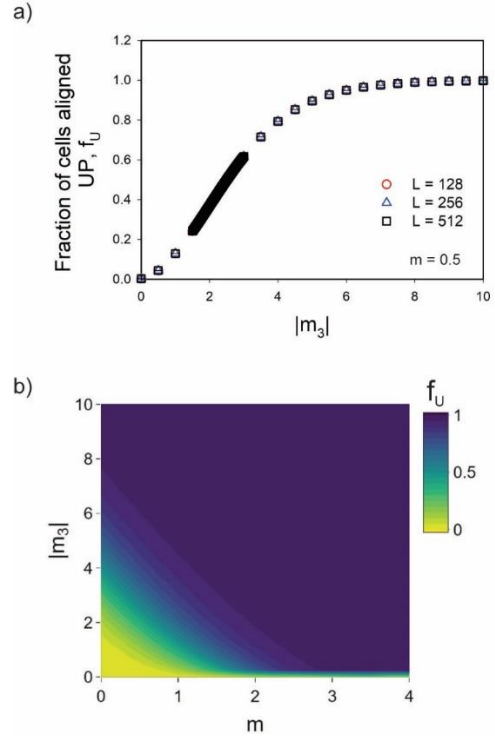

**Figure 2.** Alignment of cells in the correct (UP) direction. a) The fraction of cells aligned UP is plotted as a function of the parameter for the global cue $m_3$. The parameters for local interactions are $|m_1| = |m_2| = m = 0.5$. Data for three lattice sizes are shown. $|m_3|$ is varied at an interval of 0.5. However, in the transition region, it is varied at an interval of 0.01. b) Contour plot showing the fraction of cells correctly aligned for different combinations of $m$ and $|m_3|$. $L = 256$.

## B. Percolation transition through cluster formation

We investigated the formation of clusters of aligned cells in our model (Fig. 3). For low $|m_3|$, the system is monodispersed with mostly unaligned cells. However, with the increase in $|m_3|$, small and isolated clusters appear as more cells align in the correct direction. Beyond a threshold value of $|m_3|$, these clusters fuse to create a giant cluster of cells aligned in the UP direction. Eventually, almost all the cells in the lattice become part of this giant cluster. Such threshold behavior is typical of percolation transition.

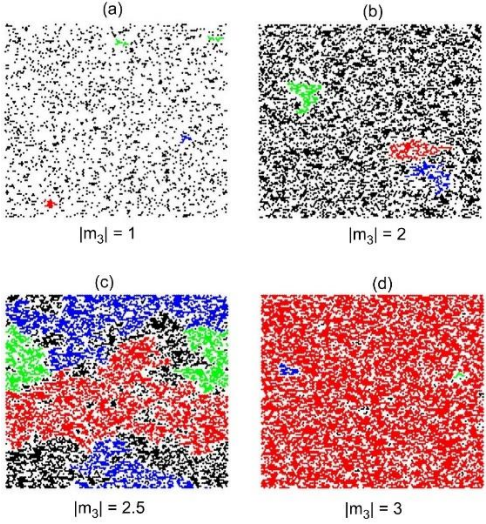

**Figure 3.** Snapshots of the lattice showing emergence of clusters for different values of $|m_3|$. The largest cluster of aligned cells is shown in red. Blue and green are the second and third largest clusters. Smaller clusters are shown in black. Unaligned cells are colored white. $L = 128$ and $m = 0.5$.

The Percolation Strength [17] captures the threshold behavior in a percolation transition. For our work, Percolation Strength is defined as the largest cluster of cells aligned in the UP direction relative to the size of the system [18]: $P = C/L^2$. Here, $C$ is the size of the largest cluster.

Fig. 4a shows the behavior of P for different lattice sizes when $m = 0.5$. Here the critical threshold for the percolation transition is somewhere between $|m_3| = 2$ and 3. $P$ is very low below this threshold but increases sharply beyond the threshold. This behavior of $P$ is a characteristic feature of continuous transition in percolation and is also observed for other values of $m$ (Fig. 4b).

Note that all cells can align in the correct direction without local interactions if the global cue is strong enough. This phenomenon is evident in the data for $m = 0$ in Fig 4b. It also shows that percolation transition appears even without local interactions.

As shown in Fig. 4b, in the presence of local interactions ($m > 0$), the percolation transition happens at lower $|m_3|$. That means when the local interactions are adequately strong, a weak global cue can align most cells in the correct direction forming a giant cluster spanning the whole system.

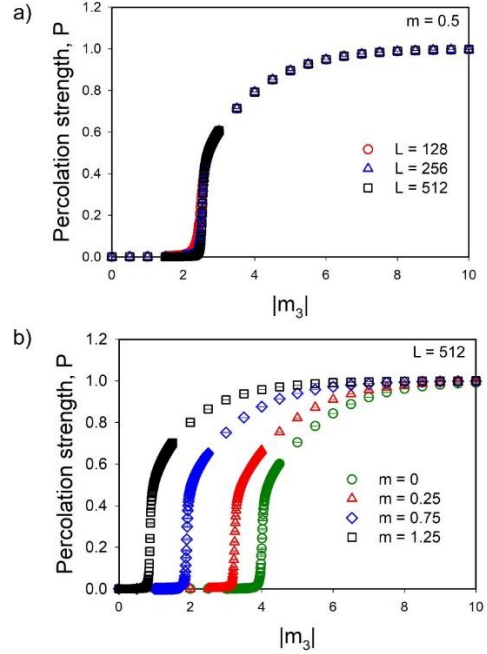

**Figure 4.** Percolation transition for different local interactions and the global cue. a) Percolation strength vs. $|m_3|$ plot for three lattice sizes with $m = 0.5$. b) Effect of $m$ on the behavior of percolation strength. Size of the lattice L = 512.

### C. Characterization of the percolation transition

We used finite-size scaling analysis to characterize the percolation transition in our model. The following scaling ansatzes are used to characterize the continuous phase transition in a finite $L \times L$ system near the percolation threshold [18].

$$P = L^{-\beta/\nu} F[(p - p_c)L^{1/\nu}] \qquad (4)$$

$$\chi = L^{\gamma/\nu} G[(p - p_c)L^{1/\nu}] \qquad (5)$$

Here $F(\cdot)$ and $G(\cdot)$ are universal functions. $P$ and $\chi$ are the percolation strength and average cluster size, respectively. The average cluster [18] size is defined as,

$$\chi = \frac{\sum_s n_s s^2}{\sum_s n_s s} \qquad (6)$$

$s$ is the size of a cluster and $n_s$ is the corresponding cluster number density. Cluster number density $n_s$ is defined as the number of clusters of size $s$ per cell.

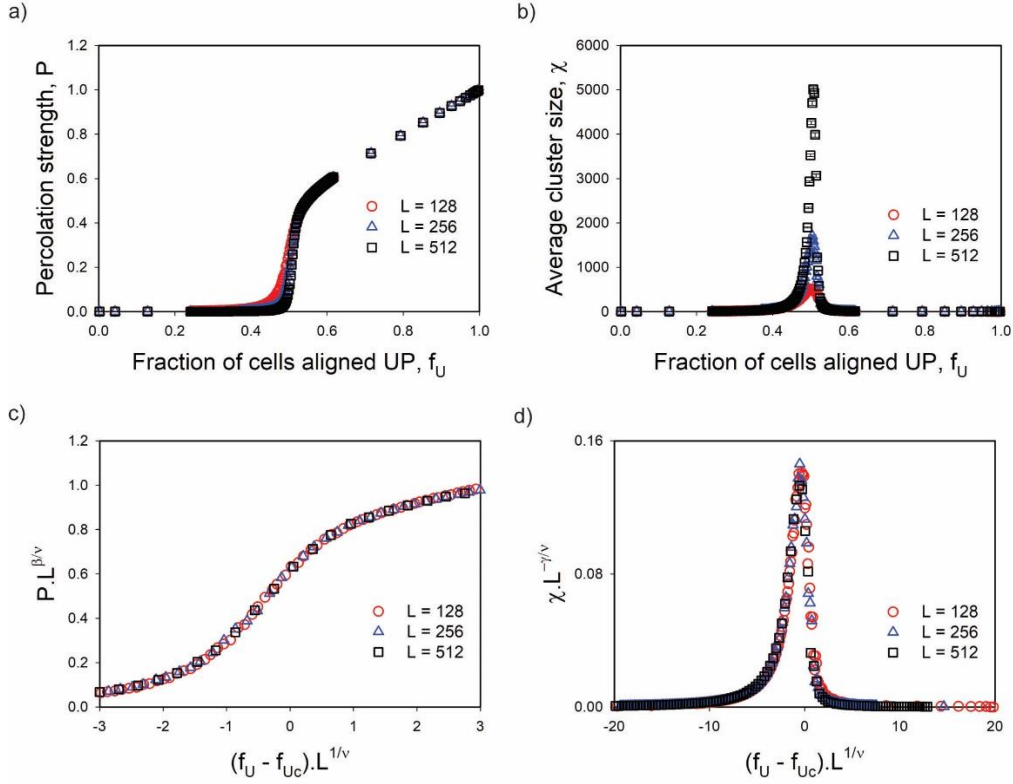

**Figure 5.** Finite-size scaling. a) and b) show the behaviors of $P$ and $\chi$ as functions of $f_U$, when $m = 0.5$ and $0 \leq |m_3| \leq 10$ for three lattice sizes. c) and d) shows the collapse of the data from three lattice sizes.

Note that the summations in Eq. (6) is over all the cluster sizes except the largest one.

In Eq. (4) and (5), $\beta$, $\nu$, and $\gamma$ are the critical exponents. The numerical values of these exponents decide the universality class of a percolation model. $p$ is the control parameter for the percolation transition and $p_c$ is the percolation threshold. In a standard 2D site percolation model, $p$ is the probability of occupancy of a site.

Our model does not have $p$ as control parameter. The control parameters for our model are $m_1$, $m_2$, and $m_3$. However, we can calculate the fraction of cells aligned in the correct direction ($f_U$) at the equilibrium for a particular combination of these parameters. So, $f_U$ of our model is equivalent to $p$ of the random site percolation model.

Using this analogy, we re-write the scaling ansatzes in terms of $f_U$ [14]:

$$P = L^{-\beta/\nu} F[(f_U - f_{Uc})L^{1/\nu}] \tag{7}$$

$$\chi = L^{\gamma/\nu} G[(f_U - f_{Uc})L^{1/\nu}] \tag{8}$$

Here $f_{Uc}$ is the percolation threshold.

Fig. 5a and 5b show the behaviors of $P$ and $\chi$ as functions of $f_U$ when $m = 0.5$ and $0 \leq |m_3| \leq 10$. The observed behaviors are as expected in continuous percolation transition. Subsequently, we estimated the critical exponents and the percolation threshold in Eq. (7) and (8), from this data set using an automatic data collapse technique [19]. The results of the data collapse are shown in Fig 5c and 5d. The estimated values of the critical exponents are: $1/\nu = 0.77(2)$, $\beta/\nu = 0.116(3)$ and $\gamma/\nu = 1.68(1)$. These estimated values are close to the critical exponents of 2D random percolation [17].

Finite-size scaling analysis was performed with data obtained from simulations with several other values of $m$, including $m = 0$. Table I shows the range of the critical exponents obtained for those simulations. The values obtained in our model for different parameters are close enough to the expected values for 2D random percolation. Therefore, we conclude that, like our previous model, the directional PCP model also belongs to the universality class of 2D random percolation.

**Table I:** Comparison of critical exponents of the PCP model with 2D random percolation

| Critical index | PCP Model (min-max) | 2D Percolation* |
|---|---|---|
| $1/\nu$ | 0.706(24) – 0.796(19) | 0.75 |
| $\beta/\nu$ | 0.104(6) – 0.124(3) | 0.104 |
| $\gamma/\nu$ | 1.55(2) – 1.70(1) | 1.785 |

\* Ref. [17]

For 2D random site percolation on a triangular lattice, the percolation threshold is 0.5 [17]. The percolation threshold ($f_{Uc}$) of our model varied from 0.5098(5) to 0.5184(5).

In our simulations, we fixed $|m_1| = |m_2| = m$ to a particular value and varied $m_3$ in a range. After the data collapse, we mapped the estimated $f_{Uc}$ to a specific value of $m_3$ using cubic spline interpolation of $f_U$ vs. $m_3$ data. For the data shown in Fig 2a and 4a ($m = 0.5$ and $0 \le |m_3| \le 10$), the estimated $f_{Uc} = 0.5112$ and this threshold corresponds to $m_3 = -2.55$.

To further confirm that our model belongs to the universality class of 2D random percolation, we checked the behavior of cluster number density at the percolation threshold. At the percolation threshold, cluster number density ($n_s$) follows power law behavior: $n_s \sim s^{-\tau}$. For 2D random percolation $\tau = 2.055$ [17].

When $m = 0.5$, the estimated percolation threshold is $m_3 = -2.55$. We checked the power-law scaling for these parameter values (Fig. 6). The estimated value of $\tau$ is 2.055(11) [$R^2 = 0.99$ and $p < 0.0001$].

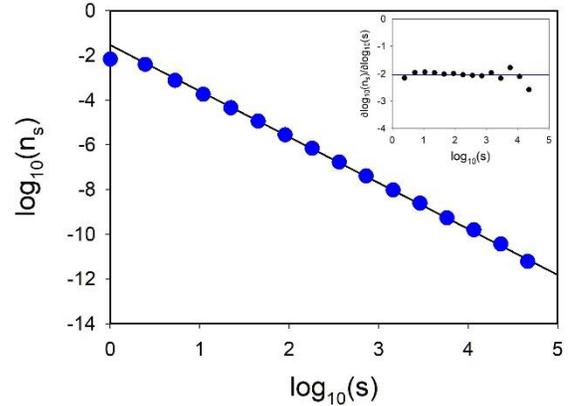

**Figure 6:** Power law scaling of cluster number density at the percolation threshold. Blue, filled circles in the double logarithmic plot show the relation between cluster number density $n_s$ and cluster size $s$. The black line is a reference line with a slope 2.055. Inset: Behavior of the local gradient. The horizontal line represents $-\tau = -2.055$. For this simulation $m = 0.5$, $m_3 = -2.55$, $L = 512$.

### D. Global cue vs. local interactions

We observed that complete alignment can happen through a strong global cue even without local interactions, and a large parameter space is available for achieving complete alignment of cells in the correct direction (Fig. 2b). These observations beg two questions. First, if complete alignment is possible through a strong global cue, why do the cells require local interactions for PCP?

Secondly, a large parameter space that allows complete alignment is somewhat unrealistic. The allowed parameter space for a cellular process should be neither very narrow nor very broad. A very narrow parameter space will not allow the system to adapt to the fluctuations in the system and the surroundings. On the other hand, a vast parameter space will negate the specificity of molecular processes.

To investigate these two questions, we introduced mutated cells in our model. These mutated cells can not sense the global cue. Therefore, the alignment of spins in these cells relies on the local interactions only. In our model, most cells are of wild type, and mutated cells are dispersed randomly in the lattice.

Fig. 7 shows the effect of mutation on the alignment of cells. As shown in Fig 7, correct alignment of all the cells in the lattice can happen even when 10% of the cells are mutated and can not sense the global cue. These mutated cells get correctly aligned by interacting with their wild-type neighbors. These would not have been achieved if the PCP relied only on global signaling. Therefore, local interactions are required to make the system robust in the face of any local disruptions in global signaling.

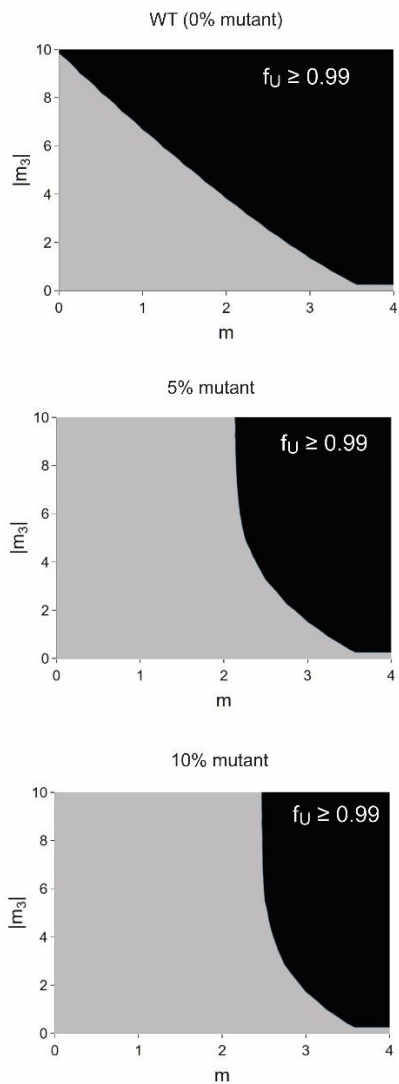

**Figure 7.** Effect of mutation on the alignment of cells. Data for three situations are shown. The $m$ vs. $|m_3|$ parameter space is divided in two zones. The black region represents the parameter space where all the cells are aligned in the correct direction ($f_U \geq 0.99$). The light grey region represents $f_U < 0.99$. The percentages of cells mutated are given above the figures. For mutated cells $m_3 = 0$. WT means wild type. $L = 256$.

Fig. 7 also shows that the parameter space that allows complete alignment (the black regions in the plots) shrinks drastically with mutation. However, most of the shrinkage happens in the region of low $m$ and high $|m_3|$. The region $m < 2$ is most affected by the mutation. This behavior is more apparent in Fig. 8.

Fig. 8 shows the sum of the square deviations of $f_U$ of a mutated model from the wild-type one for different values of $m$. For a particular value of $m$, the sum of square deviations is defined as,

$$SSD_m = \sum_{|m_3|=0}^{10} \left( f_U^{WT,|m_3|} - f_U^{mut,|m_3|} \right)^2$$

Here $f_U^{WT,|m_3|}$ is the fraction of cells aligned UP in the wild-type model for a given value of $|m_3|$. Similarly, $f_U^{mut,|m_3|}$ is the fraction of cells aligned UP in a mutated model for a given value of $|m_3|$. As shown in Fig. 8, the deviations of the mutated models are negligible when $m > 2$.

In our earlier model of PCP [14], the percolation threshold was $|m_1| = |m_2| = m = 2.1254$. Beyond this threshold, cells show a long-range correlation forming large clusters of cells aligned in the same direction. However, no global cue existed. So alignment happened in any of the six possible directions.

A similar phenomenon is happening in this directional PCP model. The only difference is that the global cue provides additional help to align in the specified direction. So for $m > 2$, the local interactions are strong enough to generate correlated alignment of cells. Even if the mutated cells do not sense the global cue, the neighboring wild-type cells aligned in the correct direction force the mutated cells to get aligned correctly.

Therefore, in a real system $m$ must be sufficiently higher than 2, so the system is insensitive to local disruptions of global cue to a certain degree. Further, when $m$ is reasonably high (say $3 \leq m \leq 4$), a weak global cue is adequate to align all cells in the correct direction, even in a mutated system (Fig 7).

Several investigators have suggested that the global cue could be transient, setting the stage for the directional segregation of molecules [20,21]. We tested this hypothesis in our simulations by setting $m_3 = 0$ after the burning phase. The extent of alignment of cells for different parameter values is shown in Fig. 9. It is amply clear that when $m$ is sufficiently higher than 2, a transient global cue is good enough to achieve complete alignment in the correct direction. Once the system has reached equilibrium, the global signal can be switched off, and the system will remain trapped in the same equilibrium.

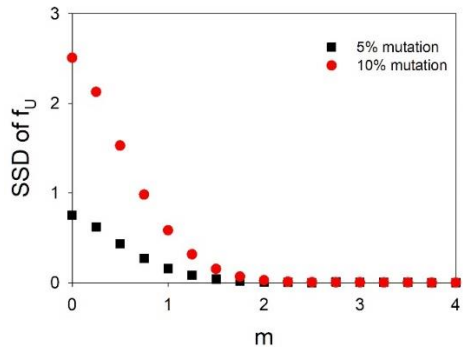

**Figure 8.** Deviation of a mutated model from the wild type. The sum of square deviations (SSD) of $f_U$ for two mutated models for different values of $m$ is shown.

From these observations, we propose that a real biological system may prefer to use a small parameter space of strong local interactions and weak global signaling to achieve collective polarization of cells in a particular direction.

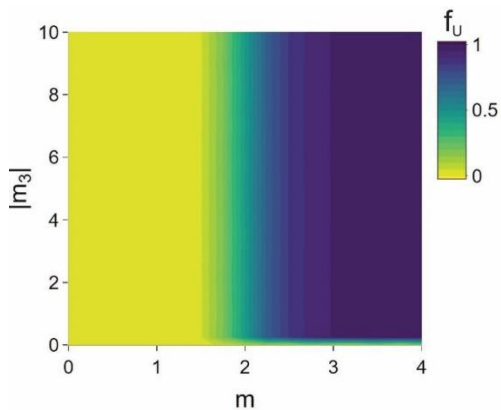

**Figure 9.** Effect of a transient global cue on the alignment of cells. Contour plot showing the fraction of cells correctly aligned for different combinations of $m$ and $|m_3|$. In these simulations, $m_3$ is used only in the burning phase and is set to zero after the burning phase. $L = 256$.

## IV. Discussion

Following the molecular mechanism of planar cell polarity in the *Drosophila* wing, we created a lattice-based spin model and investigated the emergence of system-wide collective polarization of cells in one direction. This model considers local interactions and global cues for driving the collective alignment of cells.

Earlier, we investigated a similar model with only local interactions [14]. Though the identity of the global cue in PCP in the *Drosophila* wing is still debated, it is obvious that the system needs a global cue to specify the proximal-distal direction.

The current model considers a uniform external field that acts on the spins and specifies the intended direction of polarization. This field promotes the movement of $+1$ spins to the UP direction and the opposite for $-1$ spins. In a way, this is equivalent to the movement of a charge in a uniform electric field. These preferential movements of spins are equivalent to the preferential transport of PCP-related proteins from one end of a cell to another in the *Drosophila* wing [22,23].

The global cue in PCP is usually considered a graded one. For example, Dachsous and Fourjointed show a tissue-wide gradient in expression [24]. This gradient is thought to be a global cue in PCP [12,24]. However, when the identity of the global cue is still debated, we prefer to consider a uniform field that adds the vectorial bias in the exchange of spins. That limits the number of parameters in the model.

This external field can represent the tissue-level mechanical stress originating during morphogenesis or the tissue-level expression of one or more global regulators. Whatever the global cue, that must provide a vectorial bias to the inner working of an individual cell on the wing. Therefore, the assumption of a uniform external field in a particular direction is adequate for our model.

We followed the formation of clusters of cells aligned in the UP direction. Like our previous model, this directional model of PCP also shows a percolation transition. Beyond a threshold of parameter values, cells collectively polarize in the UP direction and form large clusters of aligned cells. Eventually, these clusters fuse to create a giant cluster of aligned cells covering the whole lattice.

In our previous work, we characterized the percolation transition in terms of $|m_1| = |m_2| = m$. In this modified model, we kept $m$ constant and characterized the percolation transition in terms of the parameter of the global cue ($m_3$). We observed that this directional model of PCP belongs to the universality class of 2D random percolation.

The emergence of order through collective behavior is observed in different scales in Biology [25–27]. Ordered structures can be achieved through a global cue/ regulation, even without local interaction. Therefore, it is pertinent to ask about the relevance of local interactions in the emergence of ordered structures in a collective system [27].

In our model, all cells will be aligned in the correct direction by a strong global cue, even without local interactions. However, in the presence of appropriate local interactions, a weaker global cue can trigger the collective alignment of cells in the right direction. Such a combination of local interactions and a global cue makes the system robust against local failure of global regulation.

However, the effective regulation of collective behavior is not just a combination of local and global regulations. Rather the threshold behavior of percolation transition plays a critical role. Without the global cue, the percolation threshold of this system is at $|m_1| = |m_2| = m = 2.1254$ [14]. Beyond this threshold, cells lose their individuality and start working collectively. However, such collective alignment of cells can happen in any possible direction. Under this situation, even a weak global cue constrains the direction of alignment and nudges all the cells to align in the required direction. This behavior is evident from our experiments with mutated cells. When $m$ is sufficiently higher than this threshold value, even a 10% mutation failed to affect the collective alignment of cells in the UP direction.

Similarly, when $m$ is adequately higher than its percolation threshold, the system needs only a transient global signal until equilibrium. Once the equilibrium is reached, the collective alignment can be maintained through local interactions.

These observations indicate that to be robust against mutations in the global module or any local failure in the global cue, this system should rely on a local interaction stronger than the percolation threshold and a weak global cue. The experimental results on PCP in the *Drosophila* wing also indicate such asymmetric strengths of local and global regulations [28].

Experimental evidence proved beyond doubt that local interactions through the core PCP pathway are indispensable. A tissue-level global cue is also essential. However, we have not been able to fix the identity of that regulator. Possibly, not just one but multiple tissue-level processes during morphogenesis define the correct direction for PCP, and their regulatory strength is weaker than the strength of regulation by the local interactions. Further, such a cue could be transient, as we have observed that the global cue could be switched off once the system has reached equilibrium. A weak, transient cue is difficult to confirm through experiments.

The workings of the core PCP pathway are deciphered through experiments where some cells in the wing are mutated for specific proteins. However, such a strategy will not work to identify molecular processes of global regulation. As observed in our mutation experiment, strong local interactions will supersede any local defect. This is another impediment in the experimental identification of global regulatory processes.

Following the behavior of our model, we suggest that tissue-level intervention in both local and global signaling is required to validate a putative global regulator. Local interactions should be weakened across all cells in the tissue by intervening with the core PCP module. Such intervention should be able to cause tissue-level disturbance in the polarization of cells. Subsequently, the putative global cue should be strengthened to revert the effect of weak local interactions. Such interventions are possible through selective overexpression or repression of the expression of proteins.

# References

[1] J. P. Campanale, T. Y. Sun, and D. J. Montell, *Development and Dynamics of Cell Polarity at a Glance*, J. Cell Sci. 130, 1201 (2017).
[2] E. K. Vladar, D. Antic, and J. D. Axelrod, *Planar Cell Polarity Signaling: The Developing Cell's Compass.*, Cold Spring Harb. Perspect. Biol. 1, 1 (2009).
[3] D. Devenport, *The Cell Biology of Planar Cell Polarity*, J. Cell Biol. 207, 171 (2014).
[4] D. J. Henderson, D. A. Long, and C. H. Dean, *Planar Cell Polarity in Organ Formation*, Curr. Opin. Cell Biol. 55, 96 (2018).
[5] D. L. Shi, *Planar Cell Polarity Regulators in Asymmetric Organogenesis during Development and Disease*, J. Genet. Genomics 50, 63 (2023).
[6] E. Papakrivopoulou, D. J. Jafree, C. H. Dean, and D. A. Long, *The Biological Significance and*

*Implications of Planar Cell Polarity for Nephrology*, Front. Physiol. 12, 1 (2021).
[7] L. Li, H. Li, L. Wang, S. Wu, L. Lv, A. Tahir, X. Xiao, C. K. C. Wong, F. Sun, R. Ge, and C. Y. Cheng, *Role of Cell Polarity and Planar Cell Polarity (PCP) Proteins in Spermatogenesis*, Crit. Rev. Biochem. Mol. Biol. 55, 71 (2020).
[8] G. Wu, X. Huang, Y. Hua, and D. Mu, *Roles of Planar Cell Polarity Pathways in the Development of Neutral Tube Defects*, J. Biomed. Sci. 18, 1 (2011).
[9] M. R. Deans, *Conserved and Divergent Principles of Planar Polarity Revealed by Hair Cell Development and Function*, Front. Neurosci. 15, 1 (2021).
[10] S. M. T. W. Maung and A. Jenny, *Planar Cell Polarity in Drosophila*, Organogenesis 7, (2011).
[11] H. Strutt and D. Strutt, *Asymmetric Localisation of Planar Polarity Proteins: Mechanisms and Consequences*, Semin. Cell Dev. Biol. 20, 957 (2009).
[12] W. Y. Aw and D. Devenport, *Planar Cell Polarity: Global Inputs Establishing Cellular Asymmetry*, Curr. Opin. Cell Biol. 44, 110 (2016).
[13] B. Ewen-campen, T. Comyn, E. Vogt, N. Perrimon, B. Ewen-campen, T. Comyn, E. Vogt, and N. Perrimon, *No Evidence That Wnt Ligands Are Required for Planar Cell Polarity in Drosophila*, CellReports 32, 108121 (2020).
[14] K. Chandrasekaran and B. Bose, *Percolation in a Reduced Equilibrium Model of Planar Cell Polarity*, Phys. Rev. E 100, 032408 (2019).
[15] C. Strandkvist, *Mathematical Models for Planar Cell Polarity*, http://strandkvist.net/wp-content/uploads/2011/07/CharlotteStrandkvistCP1.pdf.
[16] M. E. J. Newman and G. T. Barkema, *Monte Carlo Methods in Statistical Physics* (Clarendon Press, 1999).
[17] D. Stauffer and A. Aharony, *Introduction to Percolation Theory* (Taylor & Francis, London, 1994).
[18] F. Radicchi and S. Fortunato, *Explosive Percolation in Scale-Free Networks*, Phys. Rev. Lett. 103, 1 (2009).
[19] O. Melchert, *AutoScale.Py - A Program for Automatic Finite-Size Scaling Analyses: A User's Guide*, arXiv:0910.5403 (2009).
[20] K. Amonlirdviman, N. A. Khare, D. R. P. Tree, W.-S. Chen, J. D. Axelrod, and C. J. Tomlin, *Mathematical Modeling of Planar Cell Polarity to Understand Domineering Nonautonomy*, Science 307, 423 (2005).
[21] J. F. Le Garrec, P. Lopez, and M. Kerszberg, *Establishment and Maintenance of Planar Epithelial Cell Polarity by Asymmetric Cadherin Bridges: A Computer Model*, Dev. Dyn. 235, 235 (2006).
[22] Y. Shimada, S. Yonemura, H. Ohkura, D. Strutt, and T. Uemura, *Polarized Transport of Frizzled along the Planar Microtubule Arrays in Drosophila Wing Epithelium*, Dev. Cell 10, 209 (2006).
[23] M. Matis, D. A. Russler-Germain, Q. Hu, C. J. Tomlin, and J. D. Axelrod, *Microtubules Provide Directional Information for Core PCP Function*, Elife 3, 1 (2014).
[24] D. Ma, C. hui Yang, H. McNeill, M. A. Simon, and J. D. Axelrod, *Fidelity in Planar Cell Polarity Signalling*, Nature 421, 543 (2003).
[25] A. Xavier da Silveira dos Santos and P. Liberali, *From Single Cells to Tissue Self-Organization*, FEBS J. 286, 1495 (2019).
[26] R. Wedlich-Söldner and T. Betz, *Self-Organization: The Fundament of Cell Biology*, Philos. Trans. R. Soc. B Biol. Sci. 373, (2018).
[27] S. Camazine, J. Deneubourg, N. R. Franks, J. Sneyd, G. Theraulaz, and E. Bonabeau, *Self-organization in Biological Systems* (Princeton University Press, 2020).
[28] A. Brittle, S. J. Warrington, H. Strutt, E. Manning, S. E. Tan, and D. Strutt, *Distinct Mechanisms of Planar Polarization by the Core and Fat-Dachsous Planar Polarity Pathways in the Drosophila Wing*, Cell Rep. 40, 111419 (2022).